\documentclass[11pt]{article}
\usepackage{paspconf}

\font\ninerm=cmr9    \font\sixrm=cmr5
  
\font\nineit=cmti9  
\font\ninesl=cmsl9
\font\ninei=cmmi9    \font\sixi=cmmi5
\skewchar\ninei='177
\font\ninesy=cmsy9  \font\sixsy=cmsy5
\skewchar\ninesy='60
\font\ninebf=cmbx9  \font\sixbf=cmbx5
\font\nineex=cmex10 scaled 833

\font\ninett=cmtt9
\def\adjustlinespace{\baselineskip=\baselineskip}
\def\ninepoint{\textfont0=\ninerm \scriptfont0=\sixrm 
                \def\rm{\fam0\ninerm}\relax
                \textfont1=\ninei \scriptfont1=\sixi 
                \def\mit{\fam1}\def\oldstyle{\fam1\ninei}\relax
                \textfont2=\ninesy \scriptfont2=\sixsy 
                \def\cal{\fam2}\relax
                \textfont3=\nineex \scriptfont3=\nineex 
                \def\it{\fam\itfam\nineit}\relax
                \textfont\itfam=\nineit
                \def\sl{\fam\slfam\ninesl}\relax
                \textfont\slfam=\ninesl
                \def\bf{\fam\bffam\ninebf}\relax
                \textfont\bffam=\ninebf \scriptfont\bffam=\sixbf
                \def\tt{\fam\ttfam\ninett}\relax
                \textfont\ttfam=\ninett
                \setbox\strutbox=\hbox{\vrule
                     hnine7pt depth2pt width0pt}\baselineskip=9pt
                \adjustlinespace
                \rm}
\font\caps=cmcsc10			   
\font\fivebmi=cmmib6
\font\sixbmi=cmmib6	\skewchar\sixbmi='177
\font\ninebmi=cmmib10 at 9pt 	\skewchar\ninebmi='177
\newfam\bmifam
\textfont\bmifam=\ninebmi
\scriptfont\bmifam=\sixbmi
\scriptscriptfont\bmifam=\fivebmi
\def\bmi{\fam\bmifam\ninebmi}
\def\b#1{{\bmi#1}}

\def\etal{{\em et~al.\ }}
\def\aa #1 #2 {A\&A, #1, #2}
\def\aas #1 #2 {A\&AS, #1, #2}
\def\acm #1 #2 {ACM-Trans Math Software, #1, #2}
\def\ada #1 #2 {Ann Astrophys, #1, #2}
\def\agabstr #1 #2 {Astr Ges Abstr Ser, #1, #2}
\def\aj #1 #2 {AJ, #1, #2}
\def\anach #1 #2 {Astr Nachr, #1, #2}
\def\apj #1 #2 {ApJ, #1, #2}
\def\apjl #1 #2 {ApJL, #1, #2}
\def\apjs #1 #2 {ApJS, #1, #2}
\def\araa #1 #2 {ARAA, #1, #2}
\def\apss #1 #2 {ApSpaceS, #1, #2}
\def\celmech #1 #2 {Cel Mech, #1, #2}
\def\esom #1 #2 {ESO Messenger, #1, #2}
\def\fundcp #1 #2 {FunCosP, #1, #2}
\def\jcp #1 #2 {J Comp Phys, #1, #2}
\def\jfm #1 #2 {J Fluid Mech, #1, #2}
\def\jmp #1 #2 {J Math Phys, #1, #2}
\def\ma #1 #2 {Mitt Astr Ges, #1, #2}
\def\mn #1 #2 {MNRAS, #1, #2}
\def\nat #1 #2 {Nat, #1, #2}
\def\obs #1 #2 {Observatory, #1, #2}
\def\pasj #1 #2 {PASJ, #1, #2}
\def\pasp #1 #2 {PASP, #1, #2}
\def\phyr #1 #2 {PhysRep, #1, #2}
\def\physd #1 #2 {Physica D, #1, #2}
\def\rpp #1 #2 {RepProgPhys, #1, #2}
\def\ssr #1 #2 {Sp Sci Rev, #1, #2}
\def\iau127#1{in de Zeeuw P.T. ed, Structure and Dynamics of 
     Elliptical Galaxies, IAU Symp.~No.~127. Reidel, Dordrecht, p.~#1}

\def\inbook#1#2#3#4#5#6{in: #1%
\if#2-%
\else%
, #2%
\fi%
\if#3-%
\else%
, ed.\ #3%
\fi%
\if#5-%
 {\if#4-%
 \else,%
   (#4)%
 \fi}%
\else%
 {\if#4-%
, (#5)%
\else%
, (#5:#4)%
\fi}%
\fi%
\if#6-%
.%
\else%
, #6.%
\fi%
}

\def\spose#1{\hbox to 0pt{#1\hss}}
\def\lta{\mathrel{\spose{\lower 3pt\hbox{$\mathchar"218$}}
     \raise 2.0pt\hbox{$\mathchar"13C$}}}
\def\gta{\mathrel{\spose{\lower 3pt\hbox{$\mathchar"218$}}
     \raise 2.0pt\hbox{$\mathchar"13E$}}}

\def\=#1{\overline{#1}}

\def\deg{^\circ}             

\def\mum{\mu{\rm m}}	     

\def\kms{{\rm\,km\,s^{-1}}}

\def\pc{{\rm\,pc}}
\def\kpc{{\rm\,kpc}}

\def\msun{{\rm\,M_\odot}}

\def\rsun{{\rm\,R_\odot}}

\def\myr{{\rm\,Myr}}

\input psfig.sty
\newif\ifpsfiles\psfilestrue

\def\rcr{R_{\rm CR}}

\def\rbar{R_{\rm bar}}

\def\OmegaP{\Omega_{\rm P}}

\def\lvplot{($l,v$) diagram}
\def\lvplots{\lvplot s}

\def\twco{$^{12}$CO}
\def\thco{$^{13}$CO}

\def\mum{\,\mu}

\def\los{{\caps LOS}}

\begin{document}

\title{Dynamics of the Galaxy}
\author{Ortwin E. Gerhard}
\affil{Astronomisches Institut der Universit\"at Basel, Switzerland, gerhard@astro.unibas.ch}

\begin{abstract}

Our Galaxy is a barred spiral.  Recent work based on the COBE NIR data
implies a small bulge--bar and a disk with a short scale-length.  The
corotation radius of the bar is in the range 3--4.5 kpc.  The stellar
density distribution beyond the end of the bar appears to be perturbed
strongly by the Galaxy's spiral arms.

Gas flow calculations in corresponding potentials provide a
qualitative explanation of many features observed in HI and CO
$lv$-diagrams. These include the 3-kpc-arm and the apparent
four--armed spiral structure between corotation and the solar radius.

The mass of NIR--luminous matter is constrained by the terminal
velocity curve, the Oort limit, and the bulge microlensing observations,
and this implies that the Milky Way has a near-maximum disk
and a dark halo with a large core radius of $\sim 15\kpc$.

However, we are still some way from a detailed quantitative model for
the large--scale dynamics of the Galaxy.  I summarize a number of
uncertainties as well as how future work might resolve them.

\end{abstract}

\section{Introduction: The Barred Milky Way}

It is now well--established that the Milky Way is barred.  This is an
important development: some long-standing questions about Galactic
dynamics are beginning to be answered in the new framework, as
discussed in later sections, and, not least, it is changing the way in
which we have to think about the Galaxy's evolutionary history.

The strongest evidence for a rotating bar in the inner Galaxy comes
from the NIR light distribution, source count observations, the atomic
and molecular gas morphology and kinematics, and the large optical
depth to microlensing.  This section gives a brief summary of this
evidence; other recent reviews concerning observations and dynamical
implications of the Galactic bar are in Gerhard (1996), Kuijken
(1996), and Morris \& Serabyn (1996).

Observations of cold gas in the inner Galaxy reveal large
non--circular motions (e.g., Burton \& Liszt 1978 [HI], Dame \etal
1987 [\twco ], Bally \etal 1988 [\thco ]). Some of the more prominent
features include the 3--kpc--arm, the 135--km/s--arm, the molecular
parallelogram or 180--pc--ring, and the high central peak in the
terminal velocity curve at $l\simeq \pm 2\deg$. Many papers in the
past have suggested that the observed kinematics are best explained by
gas motions in a barred potential (Peters 1975, Cohen \& Few 1976,
Liszt \& Burton 1980, Gerhard \& Vietri 1986, Mulder \& Liem 1986,
Binney \etal 1991, Wada \etal 1994).  In \S5 I discuss new
hydrodynamical studies of the gas flow in the barred Milky Way.

The COBE-DIRBE photometry clearly show (Weiland \etal 1994,
Freudenreich 1998, see Fig.~1) that the Galactic bulge is both
brighter and more extended in latitude at given positive longitude
than at the same negative longitude, except for a region close to the
Galactic Centre where the first effect is reversed.  These signatures
are just as expected for a triaxial bulge with its long axis in the
first quadrant (Blitz \& Spergel 1991). Detailed modelling results are
discussed in \S2.

The flux distribution for IRAS bulge sources is systematically
brighter on the $l>0$ side (Nakada \etal 1991), arguing for a
non-axisymmetric bulge. A similar signature is also found in the much
larger OGLE clump giant sample (Stanek \etal 1997). The small
intrinsic luminosity spread ($\sim 0.2 - 0.3$ mag) makes these stars
good distance indicators and allows to constrain the parameters of the
bar.  Nikolaev \& Weinberg (1997) reanalyzed the IRAS variable
population, for which distance information is available from the known
range of AGB star luminosities, and again find a barred source
distribution.

The inferred optical depth in microlensing experiments towards the
bulge (Udalski \etal 1994, Alcock \etal 1997) exceeds the values
predicted by axisymmetric mass models by a factor of $2-3$ (Kiraga \&
Paczynski 1994, Evans 1994).  This has widely been taken as further
evidence for the Galactic bar.  As discussed in \S3, this is probably
correct, but the precise interpretation of the measured event rates
needs to be clarified and the predictions from models incorporating the
bar need to be improved before quantitative conclusions can be drawn.

This article is not a comprehensive review of the `Dynamics of the
Galaxy', but is restricted to a number of topics related directly or
indirectly to the existence of the bar.  In the following I will
discuss recent work on the structure of the bulge and disk (\S2),
bulge microlensing (\S3), the stellar dynamics of the bulge \S4), the
gas flow in the Galactic disk interior to the Sun (\S5), and the mass
of the bulge and disk including consequences for the distribution
of dark matter in the Galaxy (\S6), and end with some concluding remarks
in \S7.

\section{Photometric structure of the Galactic bulge and disk}

\begin{figure}
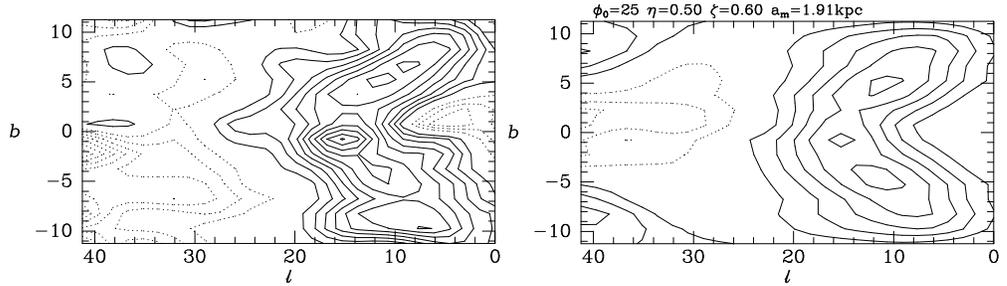

\ifpsfiles
\centerline{
\psfig{figure=diffobs.ps,width=6.5cm}
\psfig{figure=d25.ps,width=6.5cm}
}
\else
\vspace{6.5cm}
\fi
\caption{\small Left: Asymmetry map for the cleaned COBE L--band
data of Spergel \etal (1996). Contours are spaced by 0.05 mag; dotted
contours indicate larger flux on the $l<0$--side.  Right: Asymmetry
map of a model obtained by Lucy-Richardson deprojection, assuming a
bar angle of $\phi=25\deg$. From Bissantz \etal (1997).}
\end{figure}

The currently best models for the distribution of old stars in the
inner Galaxy are based on the COBE/DIRBE NIR data.  The COBE data have
complete sky coverage, and provide broad-band emission maps from
J,K,L,M in the NIR to $100\mum$ and $240\mum$ in the FIR. However,
their analysis is complicated by the relatively low spatial
resolution, the residual effects of dust absorption, and the fact that
they contain no distance information; the best results will ultimately
be obtained by combining them with other complementary data sets.
Because extinction is important towards the Galactic nuclear bulge
even at $2\,\mu$m, the first task is to correct (or `clean') the DIRBE
data for the effects of extinction.  Arendt et al.\ (1994) did this by
assuming that the dust lies in a foreground screen, while Spergel,
Malhotra \& Blitz (1996) and Freudenreich (1998) used fully
three-dimensional models of the dust distribution.

The cleaned DIRBE data confirm that the Galactic bulge is both
brighter and more extended in latitude ($b$) at positive longitudes
than at corresponding negative longitudes ($l$), except for a region
close to the Galactic Centre where the bulge is brighter at negative
$l$ (Dwek \etal 1995, Bissantz \etal 1997).  These signatures are just
as expected for a triaxial bulge with its long axis in the first
quadrant (Blitz \& Spergel 1991). They can be quantified by an
`asymmetry map' -- flux at $(l>0,b)$ divided by flux at $(-l,b)$ in
logarithmic units -- cf.\ Fig.~1 (from Bissantz \etal 1997). The
region of negative asymmetry at small $|l|$ argues for a bar rather
than a lopsided distribution as the cause of the positive asymmetry at
larger $|l|$; see also Sevenster (1997).  Most of the asymmetry signal
is in the range $0.1-0.3$ mag.

Binney \& Gerhard (1996) developed a Lucy--Richardson deprojection
algorithm to interpret these data, based on the assumption that the
Galaxy's NIR emissivity distribution is eight--fold (triaxially)
symmetric . They showed that in this case the ambiguity inherent in
the deprojection of 2d data is reduced to essentially that involved in
choosing the orientation of the bulge's symmetry planes.  This
algorithm was used by Binney, Gerhard \& Spergel (1997, hereafter
BGS97) to fit non--parametric models for the 3d emissivity $\b j(r)$
to the cleaned L--band data of Spergel \etal (1996). When the
orientation of the symmetry planes is fixed, the recovered emissivity
$\b j(r)$ appears to be essentially unique, but physical models for
the COBE bar can be found for a range of bar orientations: $15\deg
\lta \phi \lta 35\deg$, where $\phi$ measures the angle in the
Galactic plane between the bar's major axis at $l>0$ and the
Sun--centre line.  Zhao (1999) has given an illustration of the
non--uniqueness of the bar orientation in terms of the even part of
the bulge density distribution.

\begin{figure}
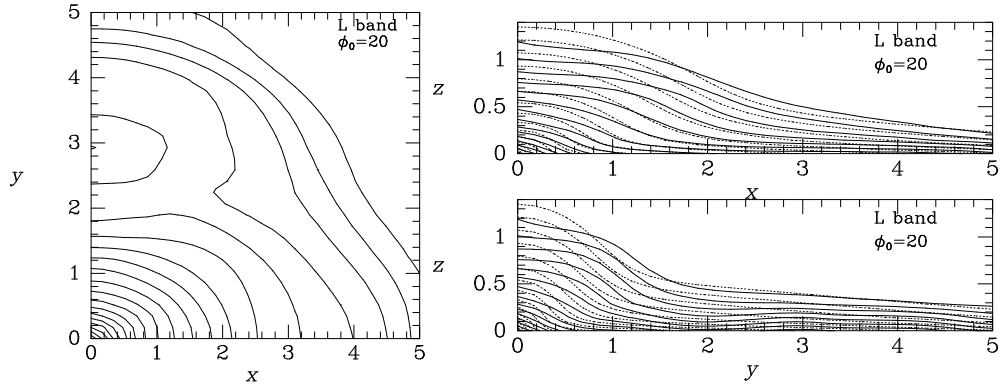

\ifpsfiles
\hbox to \hsize{
  \psfig{figure=proj20.ps,width=5.5cm}
   \hfil
  \vbox{
    \psfig{figure=twentyy.ps,width=7.5cm}
    \par
    \psfig{figure=twentyx.ps,width=7.5cm}
    \vfill
  }
}
\else
\vspace{5.5cm}
\fi
\caption{\small Symmetrized luminosity model for the inner Galaxy obtained
  by Lucy--Richardson deprojection of the cleaned COBE L--band data
  for $\phi=20\deg$. Left: Density projected along the $z$-axis, contours
  spaced by 0.1 dex. Right: Isodensity surfaces in the $zx$ and $zy$
  planes, contours spaced by 0.2 dex. Axis lengths in kpc. From Binney,
  Gerhard \& Spergel (1997).}
\end{figure}

Fig.~2 shows the deprojected luminosity distribution resulting for
$\phi=20\deg$. This shows an elongated bulge with axis ratios 10:6:4
and semi--major axis $\sim 2\kpc$, surrounded by an elliptical disk
that extends to $\sim 3.5\kpc$ on the major axis and $\sim 2\kpc$ on
the minor axis. There is a maximum in the projected NIR emissivity
$\sim 3\kpc$ down the minor axis in the Galactic plane, which is
caused by emission near the plane and is probably due to incorrectly
deprojected strong spiral arms (see below). This feature corresponds
to the ring--like structure discussed by Kent, Dame \& Fazio
(1991). An exponential fit at larger radii gives a radial disk scale
length of $R_D=2.5\kpc$.  Fig.~1 shows the asymmetry map of a similar
model to compare with the data.

Dwek \etal (1995) and Freudenreich (1998) have constructed parametric
models to match their dust--cleaned DIRBE data.  In Freudenreich's
(1998) recent work the region near the Galactic plane is excluded in
the fit (owing to uncertainties in the dust extinction) and the disk
is modelled with a central hole.  His analysis based on several NIR
wavebands gives a bulge with axial ratios 10:4:3, length $3.1\kpc$, at
a preferred orientation of around $\phi=14\deg$. The disk scale-length
outside the central hole is $R_D=2.6\kpc$.

The differences between these models perhaps give a fair measure of
the remaining uncertainties. It is likely that the dominant source of
error is still buried in the details of the dust absorption model, but
also systematic errors in constructing 3d models from 2d data have to
be explored further.  One major difference is in the structure of the
inner disk and, hence, the amount of luminosity in the inner disk
compared to the bulge.  To resolve this clearly requires fitting the
data at small $l$ and $b$ and hence a reliable dust model. Absorption
will be a lesser problem at higher latitudes where the bulge structure
is mainly determined, but the observed asymmetry signal of a few
tenths of a magnitude is not greatly larger than the estimated errors
in the extinction ($\sim 0.07 {\rm mag}$, Spergel \etal 1996). Thus it
appears worthwhile to reanalyze the DIRBE data based on more
sophisticated dust extinction models (e.g., Sodroski \etal 1997).

A different and very promising approach is based on analysing large
point-source catalogues.  Stanek \etal (1997) fitted bulge--bar models
to the apparent magnitude distributions of clump giant stars in 12
OGLE fields.  The small intrinsic luminosity spread ($\sim 0.2 - 0.3$
mag) makes these stars good distance indicators.  Stanek \etal's best
bulge model fitted to the reddening--corrected data is one with an
exponential density distribution, axial ratios 10:4:3, and a bar angle
of $\phi=20 -30 \deg$. These parameters stem from a parametric fit of
the bulge only and some contamination from the foreground disk seen in
the NIR is likely.  With potentially many more fields to come from the
OGLE II and MACHO projects this method has future potential.

Nikolaev \& Weinberg (1997) reanalyzed the IRAS variable population,
for which distance information is available from the known range of
AGB star luminosities.  This sample has a large spatial coverage near
the plane; taking into account selection effects and incompleteness,
Nikolaev \& Weinberg (1997) find a barred distribution (dominated by
the disk population) with length $\sim 3.5 \kpc$, axial ratio in the
plane 10:4, and orientation $\sim 21 \deg$.

NIR starcounts have confirmed the longitudinal asymmetry due to the
bar (Unavane \& Gilmore 1998), at a level more in accord with the
BGS97 model. L\'opez-Corredoira \etal (1997) have modelled the Two
Micron Survey Starcounts in three strips across the bulge. Their
best model is a nearly end--on ($\phi=12\deg\pm 6\deg$) triaxial
bar with axis ratios 10:5-6:3. Earlier starcount models (Robin \etal
1992, Ortiz \& L\'epine 1993) also favour a short disk scale
($R_D\simeq2.5\kpc$), but the uncertainties due to inbuilt model
assumptions are not easily quantified. Earlier claims that the
Milky Way disk could have a central hole have not been confirmed
(Kiraga, Paczynski \& Stanek 1997).

In summary, this combined work has confirmed and elucidated the barred
structure of the Galactic bulge and inner disk. While the length of
the bar is near 3--3.5 kpc and the bar angle $\phi$ with respect to
the Sun--Center line is probably around $\phi=20-25\deg$, there is
still some debate about the axis ratios (within a range
10:[4-6]:[3-4]) and the detailed density structure. A further
important conclusion with considerable consequences for Galactic mass
models (see \S 8) is that the radial scale--length of the Galactic
disk in the NIR is short, $R_D\simeq 2.5\kpc$.  Further progress will
in my view come from (i) improved models for the dust extinction, (ii)
combining the area coverage of the COBE data with the distance
information in the clump giant and other stellar samples, and (iii) a
better understanding of possible population differences between bulge
and disk.

\section{Microlensing towards the Galactic bulge}

Microlensing observations provide new constraints on the structure of
the Galactic bulge and disk. The most robust observable is the total
optical depth averaged over the observed fields. Defining
$\tau_{-6}\equiv\tau/10^{-6}$, the measured values are: for OGLE
main-sequence stars, $\tau_{-6}=3.3\pm 2.4$ ($2\sigma$) from 9 events
(Udalski \etal 1994); for MACHO main-sequence stars,
$\tau_{-6}=1.9\pm0.8$ ($2\sigma$) from 41 events (Alcock \etal 1997),
when a correction for disk contribution is removed; and
$\tau_{-6}=3.9^{+3.6}_{-2.4}$ ($2\sigma$) for MACHO's 13 clump giant
stars, averaged over $\sim 10$ square degrees centered at
$(l,b)=(2.55\deg,-3.64\deg)$ (Alcock \etal 1997).

The interpretation of these numbers is complicated. It has recently
become clear that blending or amplification bias (Alard 1997; blended
sources below the survey magnitude limit become bright enough to be
included when lensed) is important for the bulge microlensing
events. About half of all OGLE events show strong centroid shifts
(Goldberg \& Wo\'zniak 1998), hence are affected by blending. The
required correction to the optical depth has yet to be determined. The
effect is likely to be similarly important for the MACHO main sequence
stars, whereas the bright clump giant stars should be unaffected. A
proper analysis of the larger MACHO 4 yr dataset is eagerly
awaited. Another complication are the steep gradients in the bulge
expected from the NIR models; spatially resolved data would of course
be ideal for structural information.

It has long been known that axisymmetric models predict
$\tau_{-6}\simeq 1-1.2$, insufficient to explain the quoted optical
depths (Kiraga \& Paczynski 1994, Evans 1994).  Models with a nearly
end--on bar as described in \S 2 enhance $\tau$ because of the longer
average \los\ from lens to source.  The maximum effect occurs for
$\phi\simeq \arctan(b/a)$ when $\tau_{\rm bar}/\tau_{\rm axi}\simeq
(\sin2\phi)^{-1}\simeq 2$ for $\phi=15\deg$ (Zhao \& Mao 1996).  In
addition, $\tau$ increases with the mass and the length of the
bar/bulge. 

Optical depths in Baade's window in some recent bar models for the
Milky Way are $\tau_{-6}\simeq 2.2$ for $M_{\rm bulge}\simeq
2.0\,10^{10} \msun$ based on a model of the COBE data (Zhao, Spergel
\& Rich 1995), $\tau_{-6}\simeq 1.3$ for $M_{\rm bulge}\simeq
1.6\,10^{10} \msun$ based on a revised COBE model (Han \& Gould 1995),
$\tau_{-6}\simeq 1.5$ for $M_{\rm bulge}\simeq 1.5\,10^{10} \msun$
based on a model for the distribution of clump giants (Stanek \etal
1997), $\tau_{-6}\simeq 1.1$ for $M_{\rm bulge}\simeq 1.7\,10^{10}
\msun$ based on AGB stars (Nikolaev \& Weinberg 1997), and
$\tau_{-6}\simeq 0.9-1.3$ for $M_{\rm bulge}\simeq 8\,10^{9} \msun$
based on the deprojection of the COBE L--band data described in \S 2
(Bissantz \etal 1997). The latter paper calibrates the mass of the
disk and bulge by matching the predicted gas velocities to the
Galactic terminal velocity curve, whereas most others normalize their
models to the velocity dispersion in Baade's window (less secure
because of foreground disk contamination). Part of the differences
between the model predictions is due to differences in the bulge mass.

Thus even in bar models it has proven difficult to obtain the measured
optical depths. One interesting result was noted by Fux (1997): in his
N--body simulations a strong spiral arm in front of the Galactic
center could add as much as $\tau_{-6}=0.5$.  The resolution of the
large optical depth problem is clearly important.

\section{Stellar dynamics of the Galactic bulge--bar}

Do measurements of stellar kinematics in low--extinction bulge fields
support the triaxial nature of the bulge? How can they help to elucidate
its structure and to constrain the pattern speed or aspect angle?
To answer these questions requires dynamical models which link the
stellar density distribution and gravitational potential to the
kinematics of the stars observed.

Before discussing existing dynamical models, it is worthwhile to note
that Galactic kinematic observations in principle contain a very large
amount of information. (i) We can measure radial velocities as well as
proper motions. (ii) With enough stars, velocity histograms can be
obtained rather than just first and second moments. (iii) Distance
information is available through the known luminosity distribution of
the tracers. For the clump giant stars already discussed, the absolute
magnitude width corresponds to a distance uncertainty of $\sim
15\%$. The drawback of Inner Galaxy kinematic observations is, as
always, the heavy extinction. However, if the samples are
well--understood, and if it is possible only in a number of bulge
fields to obtain an approximate measurement of the distribution of
stars as a function of three velocities and \los\ distance, such data
will be highly constraining for the models, and thus worthwhile to
get!

An influential dynamical model has been that of Kent (1992). He
constructed an oblate--isotropic rotator model which successfully
reproduced essentially all the then available velocity dispersion data
in the bulge. This in itself is interesting because it suggests that
the signatures of triaxiality in these bulge fields are not
strong. Proper motion measurements in Baade's window (BW, Spaenhauer,
Jones \& Whitford 1992) confirmed that the velocity ellipsoid in BW
is nearly isotropic.  

One clear signature of triaxiality is rotation on the bulge minor
axis. Sofar the observational situation in the bulge is suggestive but
not yet clear (Tyson \& Rich 1991, Blum \etal 1994, Izumiura \etal 1995).
Another signature is a tilt of the velocity ellipsoid out of the
$(R,z)$--plane. Zhao, Spergel \& Rich (1994) analyzed a subsample of
the proper motions stars in BW with measured radial velocities and
found preliminary evidence for such a vertex deviation, perhaps
evidence for the bar.

New triaxial models for the Galactic bulge are broadly consistent with
a range of stellar kinematic data in the bulge region. Zhao (1996)
constructed a dynamical model for one of the Dwek \etal (1995) COBE
bar models augmented by a central cusp.  Fux (1997) analyzed N--body
models whose surface density is consistent with the COBE K-band data
after dereddening with a foreground screen model. Finally, H\"afner \etal
(1999) constructed a dynamical model by an extended Schwarzschild
method that reproduces the deprojected luminosity density of BGS97 and
simultaneously fits a number of kinematic constraints. The latter work
emphasises the importance of understanding the `selection function',
which describes the relative fraction of stars contributing to a given
kinematic sample, as a function of distance along the \los. By a
careful choice of sample, kinematic information can be obtained for a
fairly narrow range of distances along the \los. Such data will be the
most helpful for discriminating between models.

While the presently available stellar kinematics are thus compatible
with the COBE bar models, finding unambiguous evidence for triaxiality
or even constraining the bar parameters will require more and more
specific data.

\section{Gas dynamics in the Milky Way}

\begin{figure}
\ifpsfiles
\centerline{
  \psfig{figure=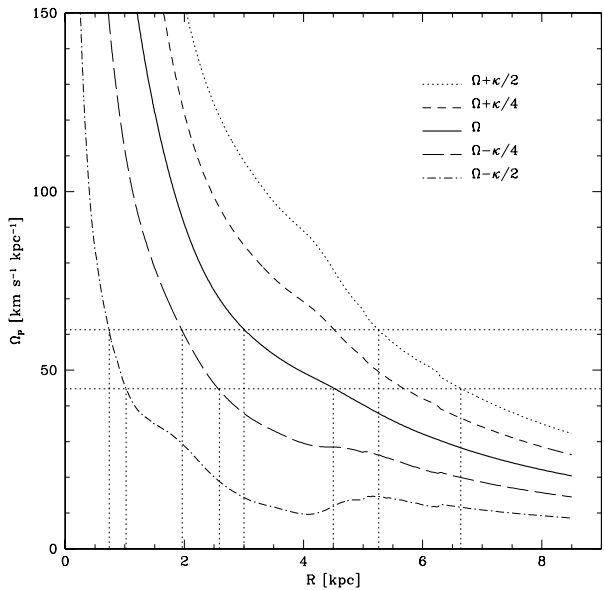,width=8.5cm}
}
\else\vskip8.5cm\fi
\caption{\small Radii of major resonances in the standard
COBE $\phi=20\deg$ bar, disk, and nucleus model of Englmaier \&
Gerhard (1999), for pattern speeds corresponding to a range of
corotation radii $R_c=3.0-4.5\kpc$.} 
\end{figure}

It has long been suggested that the non--circular and forbidden
velocities in the HI and CO \lvplots\ are the signature of a rotating
bar in the inner Galaxy (see \S 1 for references). Considerable
understanding can be reached with simple orbit--based models, because
away from resonances the gas flow is expected to approximately follow
closed ballistic orbits (e.g., Athanassoula 1992). Within such a
model, we suggested (Binney \etal 1991) that (i) the peak at $l\simeq
2\deg$ and rapid fall-off towards larger longitudes of the Galactic
terminal velocity curve can be explained in terms of the properties of
$x_1$--orbits in a barred bulge viewed nearly end--on, (ii) the dense
Galactic center clouds seen in CS are moving on $x_2$--orbits further
in, at $R\sim 100\pc$, (iii) the transition between these two flow
regimes occurs through a cusped orbit (dust lane) shock whose trace is
the parallelogram seen in the molecular gas observations of Bally
\etal (1988), and (iv) the molecular ring at $R\simeq 3.5\kpc$ is
associated with the bar's OLR.

Closed orbit models cannot take into account the effects of the strong
shocks that form in barred galaxy gas flows (e.g., Athanassoula 1992,
Englmaier \& Gerhard 1997).  Where these shocks occur, hydrodynamic
forces become important and so the flow may deviate strongly from
ballistic orbits. Also, the Binney \etal analysis was based on a
simple mass model for the inner Galaxy; now that quantitative models
can be obtained from the NIR emission of old stars, the gas flow in
the inner Galaxy can be investigated with much more confidence.
Several recent studies have thus made a fresh attack on understanding
the Galactic \lvplots, trying to model the non--circular motions in
the central $\sim 10$ degrees, the dust lane shocks, the nature of the
3-kpc-arm, the molecular ring, and the Galactic spiral arms.  These
studies have followed complementary approaches: Englmaier \& Gerhard
(1999, EG99) have modelled SPH gas flows in the deprojected COBE bulge
and disk, Fux (1999) has constructed ab initio N--body/SPH
self--consistent barred galaxy models adapted to the Milky Way, and
Weiner \& Sellwood (1999) have studied 2D fluids in a family of barred
potentials.

A primary question concerns the pattern speed $\OmegaP$, or corotation
radius $\rcr$, of the Galactic bar. We can estimate this as follows:
(i) The 3-kpc-arm is associated with non--circular velocities
$v_{nc}\simeq 50\kms$ at $l=0$. From the results of EG99, only arms
inside corotation and driven by the bar have such large $v_{nc}$. (ii)
If the rotating potential is approximately time--independent, the
inner edge of the molecular ring should be a lower limit to $\rcr$,
both if this is indeed a ring near the outer Lindblad resonance (OLR;
Schwarz 1981), and also if in reality it consists of tightly wound
spiral arms (Dame 1993, EG99).  The inner edge of the molecular ring
can be estimated from the distribution of hot cloud cores
($\rightarrow 4\kpc$, Solomon \etal 1985) and from the model of Kent
\etal (1991) for the IRT $2.4\mum$ ring ($\rightarrow
3.7\kpc$). Combining both arguments we obtain $\rcr\simeq 3.5\pm
0.5\kpc$.

An independent argument from the length of the NIR bar agrees with
this estimate. In N--body models (Sellwood \& Wilkinson 1993), and
from both direct pattern speed measurements (Merrifield \& Kuijken
1995) and gas--dynamical modelling (Athanassoula 1992), it is found
that $\rcr\simeq1.0-1.2\times \rbar$.  With $\rbar\simeq
3.2\pm0.3\kpc$ (\S2, Freudenreich 1998) this results in $\rcr\simeq
3.5\pm 0.4\kpc$ (not including possible systematic errors in locating
the end of the bar).

However, in self--consistent N--body/SPH models Fux (1999) has
observed time--dependent gas flows that never approach a
quasi--stationary configuration. The time--dependence appears to be
driven partly by a central sloshing of the bulge with amplitude of
several $100\pc$, which is not well understood, and partly by the
interaction of the bar with the outer stellar and gaseous spiral
arms. If the Milky Way is a similarly time--dependent system (see
Morris \& Serabyn 1996), argument (ii) above does not necessarily hold
because then the spiral arms in the molecular ring may penetrate into
corotation. Fitting his models to the observed \lvplot s, Fux (1999)
finds a larger $\rcr=4.0\pm0.5\kpc$.

Fig.~3 shows the radii of the major resonances for the combined range
of corotation radii $\rcr=3.0-4.5\kpc$ and a bulge, disk and nucleus
mass distribution based on the $20\deg$ model of BGS97 with constant
L-band mass--to--light ratio. The absolute scaling of the ordinate is
approximate, depending on the detailed model fit to the terminal
velocity curve. The position of the OLR is predicted to be at $5.3 -
6.7\kpc$, so is well inside the solar radius even for the slowest
pattern speed. The circular velocity at these radii from the luminous
matter only is still close to $220\kms$, so adding a dark halo near
and beyond the solar radius does not change the position of the OLR
significantly. Thus the influence of the bar on the dynamics of the
disk near the Sun should be small.

\begin{figure}
\ifpsfiles
\centerline{
  \psfig{figure=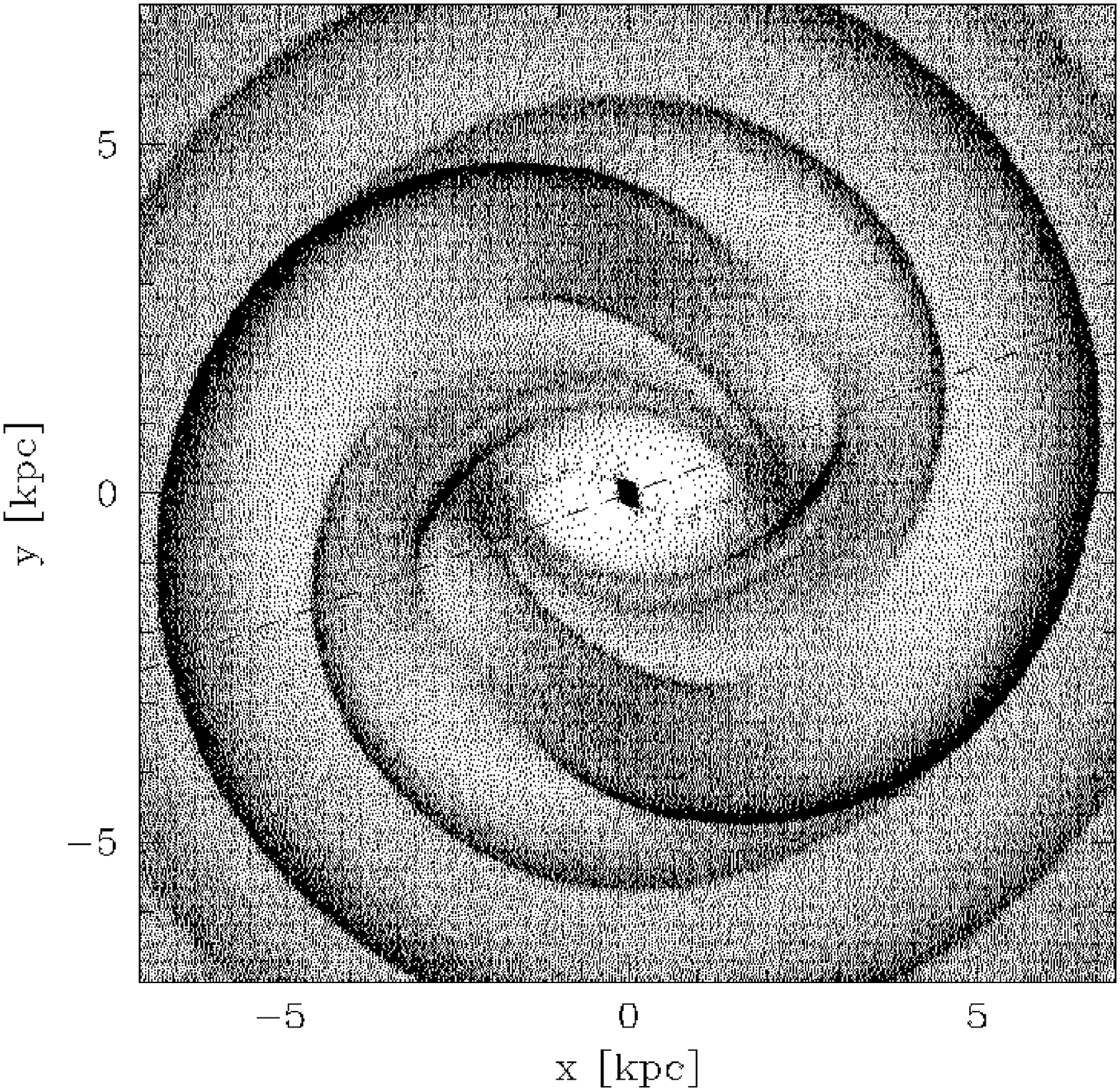,width=6.5cm}
  \psfig{figure=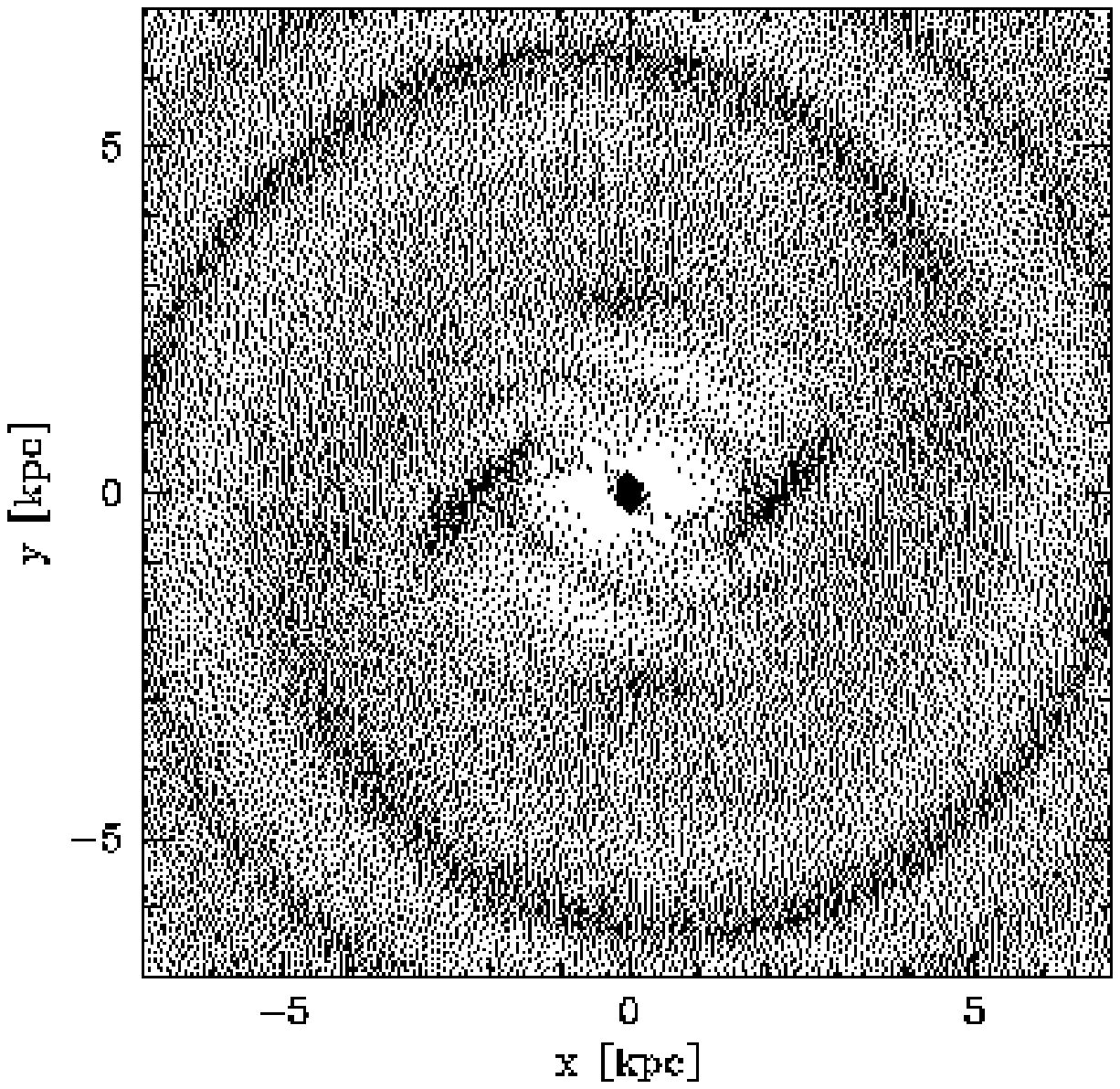,width=6.5cm}
}
\else\vskip6.5cm\fi
\caption{\small Left: Quasi--stationary gas flow in the
eightfold--symmetric COBE bulge and disk potential, deprojected for
$\phi=20\deg$, with an added nuclear cusp. Corotation is at $3.4\kpc$,
and the Sun is to the left along the dashed line.
Right: Gas flow in the same model, but at lower resolution and with
the density multipoles with $m\ne 0$ set to zero outside $3\kpc$.
From Englmaier \& Gerhard (1999).
}
\end{figure}

Is this confirmed by more detailed simulations?  Fig.~4 compares
model gas flows in the full $\phi=20 \deg$ COBE potential and in the
same model modified by axisymmetrizing the disk for radii $R\ge
3.5\kpc$. The two gas flows are dramatically different; the bar alone
is unable to drive significant spiral arm structure in the gas beyond
$R\simeq 4\kpc$ -- the structures seen in the right panel of Fig.~4
are merely gas concentrations but no real shocks; this is not a
resolution effect. Also, the two arms beginning above and below the
bar in the left panel of Fig.~4 are absent in the right panel.
The bar--only model would be in conflict with the
tangent point observations indicating the presence of strong spiral
arms at $l=-50\deg, -30\deg, 30\deg, 50\deg$, and the abundance of
warm CO clouds found near $l=25-30\deg$ (Solomon, Sanders \& Rivolo
1985) indicating a spiral arm shock in this region. In the full COBE
model, the spiral arms beyond $R\simeq 4\kpc$ are driven by the mass
corresponding to the NIR luminosity concentrations at $\gta 3\kpc$
down the bar minor axis (see \S2).  The conclusion is therefore that
these concentrations cannot just be caused by luminous supergiant
stars, but must contain significant mass.  Most likely, they are due
to incorrectly (because of the employed symmetry assumption)
deprojected spiral arms. If this interpretation is correct, the
dynamics between the bar's corotation radius and the Sun is dominated
by the Galaxy's spiral arms. These might well have a pattern speed of
their own (Sellwood \& Wilkinson 1993), and then they and not the bar
would dominate resonances near the Sun. 

Outside $\rcr$ the induced non--circular motions are fairly small.
Fig.~5 shows predicted terminal velocity curves from several gas flow
models in the gravitational potential of the COBE bar and disk, to
which in some cases a dark halo potential has been added. In these
models the bar rotates with a pattern speed corresponding to
$\rcr\simeq 3.4\kpc$ (see EG99).  Overall, the models match the
observed HI and CO terminal velocities rather well. The two main
discrepancies are in the central peak, where there are both resolution
problems and uncertainties in the potential, and in a region around
$l\simeq -20\deg$, where the disk potential from the deprojected NIR
data may be seriously in error (see \S2). Out to $R\gta 5\kpc$,
where the dark halo begins to contribute, the observed terminal
velocity curve is consistent with the radial distribution of old stellar
mass inferred from the COBE data.

These gas flow models give rise to \lvplot s which qualitatively
reproduce a number of features seen in the observed \lvplot s (see
EG99 and the paper by Englmaier \& Gerhard in these proceedings).  The
four--armed spiral structure outside corotation (see Fig.~4) is
quantitatively consistent with the observed directions to the five
main spiral arm tangents at $|l|\le 60\deg$. The 3-kpc-arm is
identified with one of the model arms emanating from the ends of the
bar and extending into the corotation region (below the bar in
Fig.~4). The cusped--orbit shock transition channels the gas onto an
inner disk on $x_2$--orbits which in the simulations of EG99 has
radius $\sim 150\pc$ and rotation velocities $\sim 100\kms$,
approximately as inferred from CS observations (Bally \etal 1988,
Binney 1994). The models fail, however, in accounting for the
magnitude and extent in longitude of the forbidden velocities seen in
low--intensity HI observations.  Weiner \& Sellwood (1999) have
specifically modelled these HI data, and find that this problem can be
resolved if the Galactic bar is inclined by a larger angle, $\phi\gta
30\deg$, as seen from the Sun. The models of Fux (1999) show, however,
that extended forbidden velocity regions can occur also for smaller
bar angles if the model is asymmetric and no longer
quasi--stationary. Thus there may not be a conflict with the smaller
$\phi$ favoured by the clump giant distribution and the large
microlensing optical depth.

The non--stationary gas flow models in the N--body and SPH simulations
of Fux (1999; see also his paper in these proceedings) give the best
match sofar to several features in the observed \lvplot s: the 3--kpc
arm, the connecting arm, the $135\kms$--arm, and the molecular
ring. They fail in that there is no spiral arm corresponding to the
$l=50\deg$ tangent, and they assert that the inner disk on
$x_2$--orbits rotating at $\sim 200\kms$ extends to $R\simeq 1\kpc$,
distinct from the dense clouds seen in CS which extend to $R\simeq
2\deg$ and appear to rotate with $\sim 100\kms$ on presumably
$x_2$-orbits (Binney 1994). The issue of a large $x_2$-disk is also
relevant for interpreting the kinematics of the bar--driven inner
arms.  The interpretation of the 3--kpc arm as one of the lateral arms
emanating from the end of the bar is similar in the models of EG99 and
Fux (1999), although the morphology of the transition from this arm
into the dust lane shock is somewhat different, probably due to
differences in the gravitational potential and differences between
quasi--stationary and time--dependent flows.

\begin{figure}
\ifpsfiles
\centerline{
  \psfig{figure=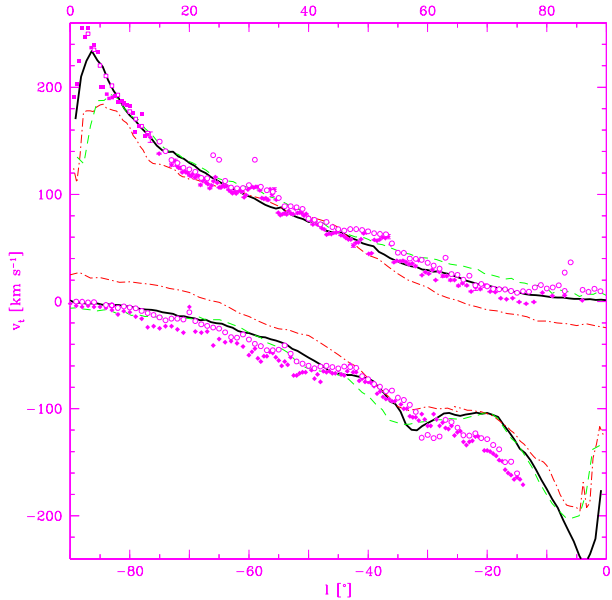,width=8.5cm}
}
\else\vskip8.5cm\fi
\caption{\small Terminal velocity curves of several gas flow models
scaled to the observed HI and CO terminal velocities (symbols; see
EG99 for references). Models are similar to that in Fig.~4:
without halo (dot--dashed); including a dark halo with asymptotic circular
speed $v_c=208\kms$, highest resolution model (full line); including
a halo with $v_c=220\kms$ (dashed line).
}
\end{figure}

Sevenster (1997) argues that in reality the 3--kpc arm is part of an
inner ring surrounding the bar, such as observed in some other
galaxies. Her strongest observational result supporting this
interpretation is (in my view) the alignment of a small group of
100--300 Myr old OH/IR stars coincident with the 3--kpc arm in the HI
\lvplot. The rotation period at 3 kpc is $\lta 100 \myr$, so these
stars appear to have remained near their birth sites for 1--3 rotation
periods. Much depends on the precise ages of these stars, and on the
pattern speed of the 3--kpc arm. From the above this will be similar
to the bar's pattern speed.  Stars at 3 kpc radius then leave the arm
on a time--scale $P_{\rm rel} = 2\pi/[\Omega(3 \kpc)-\Omega_p] \simeq
2\pi/\Omega(3 \kpc) \times [1- 3\kpc/\rcr]^{-1}$ (for flat rotation
curve). With $\rcr=[4.5\kpc, 3.5\kpc$] $P_{\rm rel}$ is $[3\times,
7\times]$ the orbital period, and for $\rcr=3\kpc$ these stars would
not leave the pattern at all.

The now existing models for the Galactic gas flow certainly do not yet
provide a quantitative fit to all the major features in the observed
\lvplots. However, with their help many of these features are
beginning to make sense in the framework of a barred Milky Way. It is
encouraging that two approaches, one starting from the COBE
observations, the other from the evolution of self--consistent N--body
systems, seem to converge to a similar picture when constrained by the
HI and and CO \lvplots: a corotation radius about half way between the
Sun and the Galactic Center, a strong, probably 4--armed spiral
pattern between $\rcr$ and the solar radius, and non--circular gas
motions inside $\rcr$ which take the gas along the inner arms near the
ends of the bar through the dust--lane shocks to the nuclear disk on
$x_2$--orbits.

\section{The mass of the bulge and disk, the Oort limit, and the \\
distribution of dark matter in the Galaxy}

The observed terminal velocity curve (Fig.~5) is consistent with a
maximum NIR disk (and bulge) model out to $|l|\simeq 45\deg$ or
$R\simeq 5.5\kpc\simeq 2R_D$, for an exponential scale length of the
disk of $R_D\simeq 2.5\kpc$ (see \S2). This maximum disk model
predicts a surface mass density of $\Sigma_\odot = 44-49 \msun/\pc^2$
near the Sun at $\rsun=8\kpc$ for a local circular velocity of
$v_{c,\odot}=208-220\kms$. The local surface density of `identified
matter' is $48\pm 8 \msun/\pc^2$, while a combined constraint from the
total surface density within $z=\pm1.1\kpc$ and the rotation curve is
$48\pm 9 \msun/\pc^2$ (Kuijken \& Gilmore 1991). Both are consistent
also with the K--giant analysis of Flynn \& Fuchs (1994).  That these
numbers approximately agree lends support to the conclusion that the
Galaxy indeed has a near--maximal disk. Similarly, the high
microlensing optical depth to the bulge argues for a high bulge and
disk mass, but a quantitative argument is presently difficult to make
(see \S3). Compared to earlier analyses, the main difference is the
short disk scale--length (see also Sackett 1997) and the extra light
around 3 kpc (see BGS97) -- the Sun is well beyond the maximum in the
rotation curve from the NIR luminous matter only.

The mass of the NIR disk inferred from the terminal velocities must
include some but may not include all of the mass in the Milky Way's
gas disk (local surface density $\sim 10\msun/\pc^2$; see Dame 1993)
and thick disk (local surface density $\sim 9\msun/\pc^2$ and scale
length $R_{\rm th}\simeq 4.5\kpc$, Beers \& Sommer--Larsen 1995, Ojha
\etal 1996), because the radial density distributions of these
components are different from the old NIR disk. For the following I
have assumed that the total visible mass in the Galaxy is given by the
mass of the COBE NIR bulge and disk and the thick disk only, assuming
that the gas disk is already accounted for by the fitted mass in the
NIR disk. Since the HI distribution has a larger scale--length than
the stars, this approximation might slightly overestimate the final
halo contribution.

Computing the gravitational forces from these two components, and
fitting a cored spherical halo so as to make the Galaxy's rotation
curve approximately flat at $R\gta 6\kpc$ with $v_c=220\kms$, results
in a halo core radius $R_c\simeq 15\kpc$. Integrating the surface
density of this halo between $z=\pm1.1\kpc$ gives $\Sigma_{h,1.1}=17
\msun/\pc^2$.  Adding this to the surface density of the old NIR disk
and the thick disk, the total is $\Sigma_{\rm NIR}+\Sigma_{\rm
th}+\Sigma_{h,1.1} = 70-75 \msun/\pc^2$, whereas the measured total
$\Sigma_{1.1} = 71\pm 6 \msun/\pc^2$ (Kuijken \& Gilmore 1991).

This very good agreement is clearly better than one expects, given the
various uncertainties in both numbers, but it shows that a maximum NIR
disk model and a near--spherical halo with a large core radius provide
a natural explanation for both the observed terminal velocity curve
and the measured surface density near the Sun. Note that for smaller
values of the Galaxy's asymptotic rotation velocity, the required
amount of halo would be reduced; for $v_c=180\kms$ the terminal
velocity curve can be fit within the errors without any added halo (at
the cost of a falling rotation curve). A more detailed analysis is
clearly worthwhile.

\section{Concluding remarks}

The work reviewed here confirms that the Galaxy contains a central
bar, seen most clearly in the NIR, the source counts and the gas
kinematic observations.  The corotation radius of the bar is about
half-way between the Sun and the Galactic Center.  There is still some
uncertainty about the orientation (most likely $\phi=20-25\deg$, but
possibly $\phi=15-35\deg$) and axial ratios (roughly 3:2:1) of the
bar, and about the relative importance of the disk in the central few
kpc. The short disk scale--length in the NIR is important; combined
with the Galactic terminal velocity curve and the Oort limit it
implies a near--maximal disk and a fairly large halo core radius.

Is there a Galactic bulge besides the bar?  This has not been
conclusively answered. While the clump giant stars appear to belong to
a strongly barred component, the NIR--luminosity could be less
strongly barred, and the RR Lyrae stars, presumably part of the
stellar halo, appear to be unbarred (Alcock \etal 1998). In external
galaxies, kpc--scale central bars and nuclear bulges may coincide
(M\"ollenhoff, private communication; see also Seigar \& James
1998). In M94, the vertical velocity dispersion in the nuclear region
is higher than that in the bar (M\"ollenhoff \etal 1995); this
suggests that in the Milky Way the question might be answered by
studying proper motions as a function of age and metallicity in low
latitude bulge fields.

\acknowledgments

This work was supported by the Swiss NSF grant 20-50676.97.

\end{document}